\documentclass[amsmath,amssymb,preprint]{revtex4}
\usepackage{graphicx}
\usepackage{dcolumn}
\usepackage{bm}

\begin{document}

\title{Electric Field Control of the LaAlO$_{3}$/SrTiO$_{3}$ Interface Ground State}
\author{A.D. Caviglia$^{1}$, S. Gariglio$^{1}$, N. Reyren$^{1}$, D. Jaccard$^{1}$, T. Schneider$^{2}$, M. Gabay$^{3}$, S. Thiel$^{4}$, G. Hammerl$^{4}$, J. Mannhart$^{4}$, J.-M. Triscone$^{1}$}
\affiliation{ $^{1}$D\'epartement de Physique de la Mati\`ere Condens\'ee, University of Geneva, 24 Quai Ernest-Ansermet, 1211 Gen\`eve 4, Switzerland}
\affiliation{ $^{2}$Physikinstitut, University of Zurich, Winterthurerstrasse 190, 8057 Zurich, Switzerland}
\affiliation{ $^{3}$Laboratoire de Physique des Solides, Bat 510, Universit\'e Paris-Sud 11, Centre d'Orsay, 91405 Orsay Cedex, France}
\affiliation{ $^{4}$Experimental Physics VI, Center for Electronic Correlations and Magnetism, Institute of Physics, University of Augsburg, D-86135 Augsburg, Germany}
%\email{Andrea.Caviglia@physics.unige.ch}
%\homepage{http://dpmc.unige.ch/gr_triscone/}
\date{\today}

\maketitle

\textbf{Interfaces between complex  oxides are  emerging as one of
the most interesting playgrounds in condensed matter physics
\cite{News-Staff:2007fj}. In this special setting, in which
translational symmetry is artificially broken, a variety of novel
electronic phases can be promoted \cite{Okamoto:2004kq}. Theoretical studies predict complex phase diagrams and suggest the key role of the carrier density in determining the systems ground states. A particularly fascinating system is the interface between the insulators LaAlO$_{3}$ and SrTiO$_{3}$, which displays conductivity with high mobility \cite{Ohtomo:2004yq}. Recently two possible ground states have been experimentally identified: a
magnetic state \cite{Brinkman:2007zr} and a two dimensional (2D) superconducting condensate \cite{N.Reyren08312007}. In this Letter we use the electric field effect to explore the phase diagram of the system. The electrostatic tuning of the carrier density allows an on/off switching of superconductivity and drives a quantum phase transition (QPT) \cite{sach, RevModPhys.69.315, lohneysen:1015} between a 2D
superconducting state and an insulating state (2D-QSI). Analyses of the magnetotransport properties in the insulating state are consistent with weak localisation and do not provide evidence for magnetism. The electric field control of superconductivity demonstrated here opens the way to the development of novel mesoscopic superconducting circuits.}

Since the discovery of conductivity at the LaAlO$_{3}$/SrTiO$_{3}$ interface,
one of the main challenges has been the identification of the source
of charge carriers. Despite an animated research effort
\cite{willmott:155502, siemons:196802, herranz:216803,
kalabukhov:121404} there is not yet a general consensus on the
nature of the doping mechanism. Depending on the growth conditions, the doping can be related to the polar
nature of the LaAlO$_{3}$ atomic planes \cite{Nakagawa:2006ys} (the
polar catastrophe scenario or ``intrinsic doping"), the creation of
oxygen defects during the samples growth \cite{herranz:216803,
siemons:196802, kalabukhov:121404} or interdiffusion phenomena
\cite{willmott:155502} (``extrinsic doping"). It is generally accepted
that the conduction observed in samples grown at low oxygen pressure
($\lesssim10^{-6}$\,mbar) is dominated by oxygen defects and extends deeply into the substrate,
while for samples grown at higher pressure ($\gtrsim10^{-5}$\,mbar) the conduction is confined at the interface \cite{Basletic:2008ee}. Both
superconductivity \cite{N.Reyren08312007} and unusual
magnetotransport properties, attributed to magnetism
\cite{Brinkman:2007zr}, have recently been observed in samples grown in the high pressure regime. These results  suggest that the ground state of the system
might be sensitive to small changes in carrier concentration and/or
to the amount of disorder. To resolve this issue and investigate the
phase diagram of the system, electrostatic doping appears to be an ideal technique, because it allows the
tuning of the carrier density while preserving the oxygen concentration and disorder landscape \cite{ahn:1185,Ahn:2003rr}. Recent reports indicate that an
electric field can effectively modulate the transport
properties of the LaAlO$_{3}$/SrTiO$_{3}$
interface \cite{S.Thiel09292006}.

\begin{figure}
\includegraphics[scale=0.3]{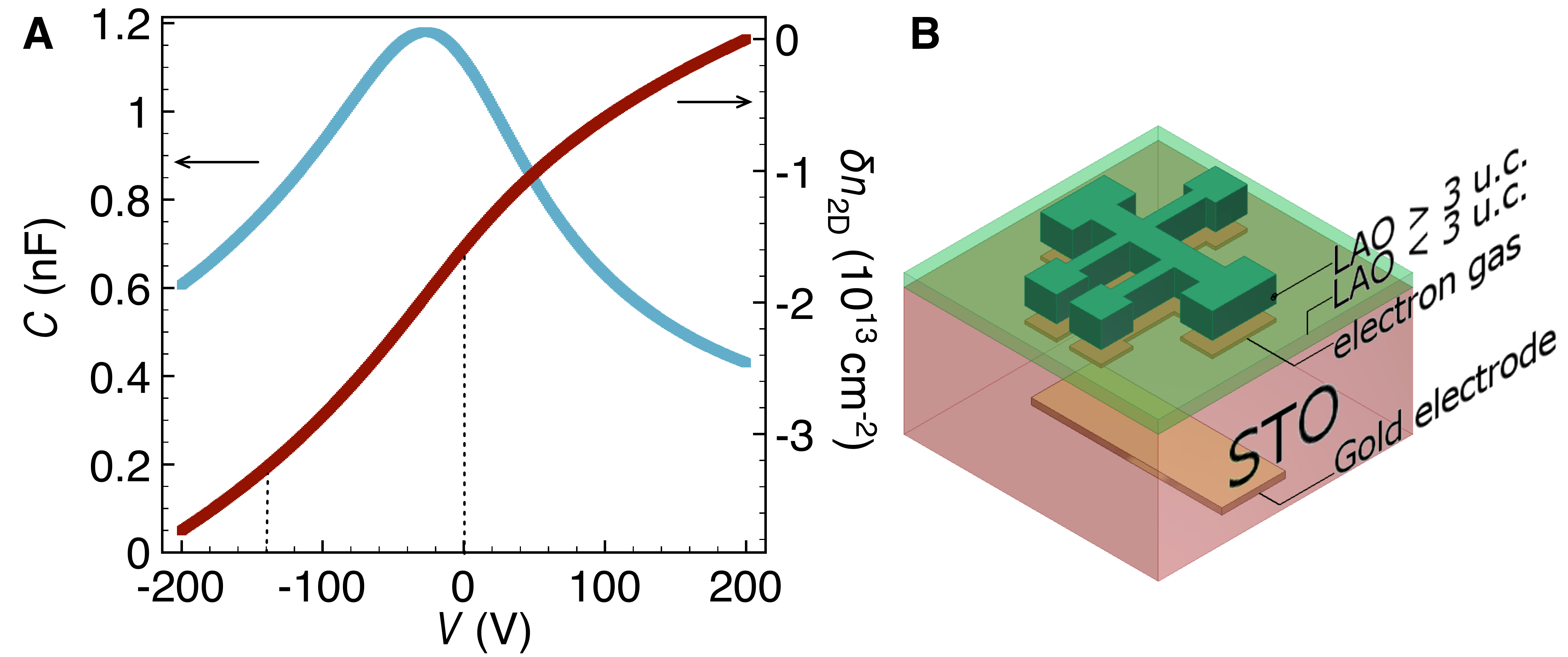}
\caption{\label{fig:dt} Dielectric characterisation of the field effect device. (A) Dielectric tunability of the differential capacitance ($C$ versus $V$) measured on a device at 1.5\,K (blue). Change in the 2D carrier concentration as a function of gate voltage calculated using eq. \ref{eq:cv} with $V_{1}= 200$\,V and $V_{2} = V$ (red). The dashed lines indicate the region where the quantum critical behaviour is observed. Note that in this region $\delta n_{2D}\sim \delta V$. (B) Schematic view of a field effect device.}
\end{figure}

LaAlO$_{3}$/SrTiO$_{3}$ superconducting interfaces and field effect devices were prepared as described in the methods section. In a standard field  effect device, an electric field is applied between a metallic gate and a conducting channel across a dielectric. The 0.5\,mm thick SrTiO$_{3}$ substrate was used as the dielectric since it is characterised at low temperatures by a large dielectric constant. The metallic gate is a gold film sputtered opposite to the channel area onto the backside of the substrate. In this
configuration the electric field modulates the concentration of
carriers in the interface conducting channel. A sketch of the field
effect device is shown in Fig. \ref{fig:dt}B.
To quantify the carrier density modulation resulting from the field effect, the electric field dependence of the SrTiO$_{3}$ permeability needs to be taken into account \cite{matthey:3758}. We therefore measured the differential capacitance $C(V)=\text{d}Q(V)/\text{d}V$ of
the device as a function of the applied gate voltage $V$, $Q(V)$ being the induced charge \footnote{$C(V)$ is measured with an ac technique applying a $\text{d}V = 1$\,V.}. It is known
that charge trapping in SrTiO$_{3}$ can occur and may cause the
appearance of hysteresis in the $C(V)$ measurements
\cite{PhysRevB.49.12095}. Indeed the capacitance was found to depend on the voltage sweep history. However, the devices present reversible and reproducible $C(V)$ characteristics when the field is
first ramped to the highest positive voltage \footnote{Note
that in our experimental procedure the 0\,V state does not
correspond to the as grown state.}. Following this
experimental procedure, the field induced
modulation of charge density $\delta n_{2D}$ between gate voltages
$V_{1}$ and $V_{2}$ can be evaluated using the relation
\begin{equation}\label{eq:cv}
\delta n_{2D} = \frac{1}{Se}\int_{V_{1}}^{V_{2}}C(V)\text{d}V
\end{equation}
where $S$ is the area of the gate electrode and $e$ is the elementary charge. The $C(V)$ characteristic of a 9\,u.c. device and the corresponding modulation of carrier density are presented in Fig.\,\ref{fig:dt}A. The initial carrier concentration $n_{2D}\simeq4.5\cdot 10^{13}$\,cm$^{2}$ has been measured using the Hall effect \footnote{The carrier density is extracted from Hall effect data at 100\,K.}. The maximum modulation of the carrier density that was achieved is remarkably close to the total number of free carriers present in the system, indicating that the electric field effect is an excellent tool to probe its phase diagram. On the same sample the temperature dependence of the sheet resistance $R_{\text{sheet}}$ has been measured for various gate voltages $V$.

\begin{figure}
\includegraphics[scale=0.3]{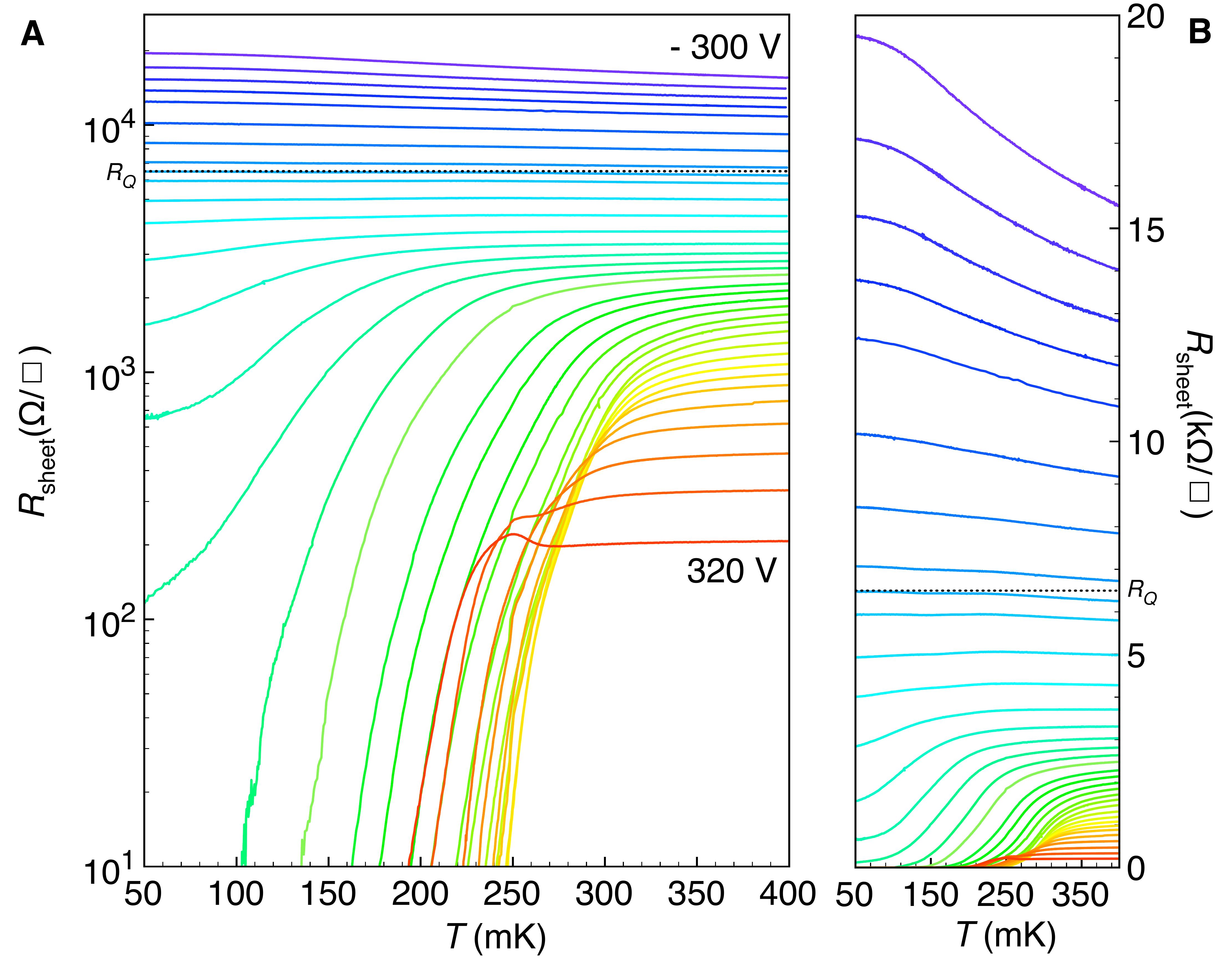}
\caption{\label{fig:rt} Field effect modulation of the transport properties. (A) Measured sheet resistance as a function of temperature for gate voltages varying in 20 V steps between -300 V and 320 V, plotted on a semi-logarithmic scale. (B) The same data plotted on a linear resistance scale.}
\end{figure}

Fig. \ref{fig:rt}A shows the sheet resistance versus temperature for
applied gate voltages between -300\,V and 320\,V; Fig. \ref{fig:rt}B
displays the same data on a linear sheet resistance scale. This behaviour has been observed in several samples. A variation of the gate voltage induces a large modulation of the normal state resistance, which changes by two orders of magnitude, and a remarkable tuning of the superconducting critical
temperature. For large negative voltages,
corresponding to the smallest accessible electron densities, the
sheet resistance increases as the temperature is decreased, suggesting an insulating (conductance $G \rightarrow 0$ as $T\rightarrow 0$) ground state. As the electron density is increased the system becomes a superconductor. The superconducting to insulating ground state transition occurs at a critical sheet resistance $R_{c} \approx 4.5$\,k$\Omega/\square$, close to the quantum of resistance for charge $2e$ bosons, $R_{Q} = h/4e^{2} \approx $ 6.45\,k$\Omega$. A further increase in the electron density produces first a rise of the critical temperature to a maximum of $\sim310$\,mK. For larger voltages, the critical temperature decreases again, providing evidence for an overdoped regime. These measurements thus reveal the existence of a quantum phase transition between a superconducting and an insulating phase at the LaAlO$_{3}$/SrTiO$_{3}$ interface and demonstrate that the ground state of the system depends on its carrier density.

\begin{figure}
\includegraphics[scale=0.3]{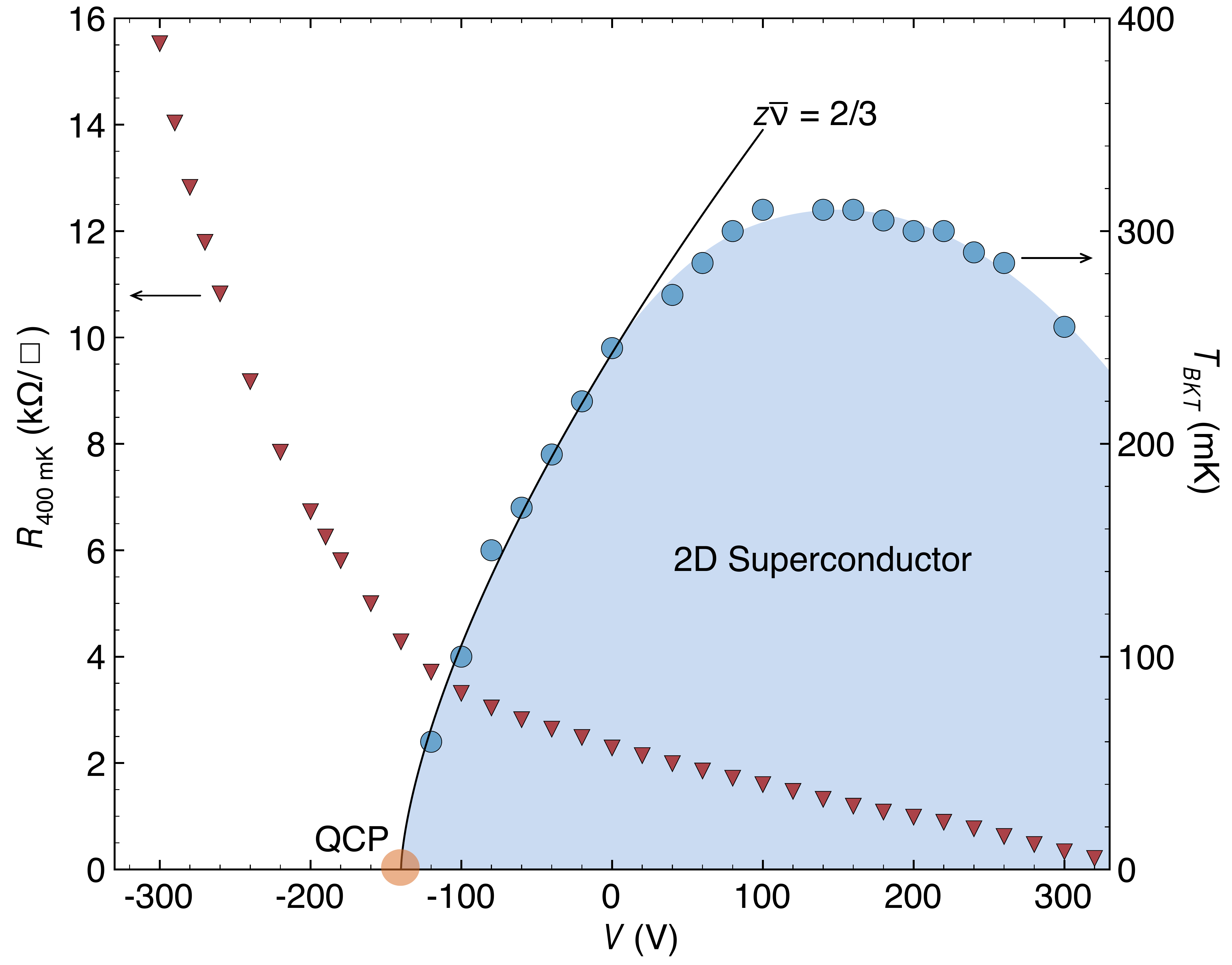}
\caption{\label{fig:vgtbkt} Electronic phase diagram of the LaAlO$_{3}$/SrTiO$_{3}$ interface. Critical temperature $T_{BKT}$ (right axis, blue dots) versus gate voltage, revealing the superconducting region of the phase diagram. The solid line describes the approach to the quantum critical point using the scaling relation $T_{BKT}\propto (V-V_{c})^{z\bar{\nu}}$, with $z\bar{\nu}=2/3$. Normal state resistance, measured at 400\,mK (left axis, red triangles) as a function of gate voltage.}
\end{figure}

Close to the critical point, the sheet resistance measured at 400 mK shows a remarkable phenomenon. As can be seen in Fig. \ref{fig:vgtbkt}, approaching the critical point from the
superconducting region of the phase diagram, a linear dependence of
the sheet resistance as a function of the applied voltage is observed. Once the
critical point has been crossed, however, a further reduction of
carrier concentration produces a much steeper variation of
resistance.

In order to establish the critical temperature versus carrier density phase diagram, a criterion to define the critical temperature for each gate voltage is needed. It was shown that the superconducting transitions are consistent with the Berezinskii-Kosterlitz-Thouless (BKT) behaviour expected for a 2D system \cite{N.Reyren08312007}. According to the BKT model, above the critical temperature
$T_{BKT}$, the temperature dependence of the resistance is expected to be
\begin{equation}
R\propto\exp\left({-\frac{b_{R}}{(T-T_{BKT})^{1/2}}}\right)
\end{equation}
where $b_{R}$ is a parameter related to the vortex properties \footnote{An analysis of the BKT transitions and finite size effects will be presented in another publication.}.
This approach allows
$T_{BKT}$  to be extracted for each applied gate voltage and  the phase
diagram to be mapped. The result is shown in Fig. \ref{fig:vgtbkt}.
Reducing the carrier concentration from the largest doping level
($V=320$\,V),  $T_{BKT}$ first increases, reaches a maximum at
around 310\,mK and then decreases to zero. This critical line
ends at $V_{c} \simeq -140$\,V, where the system undergoes a QPT.
\begin{figure}
\includegraphics[scale=0.3]{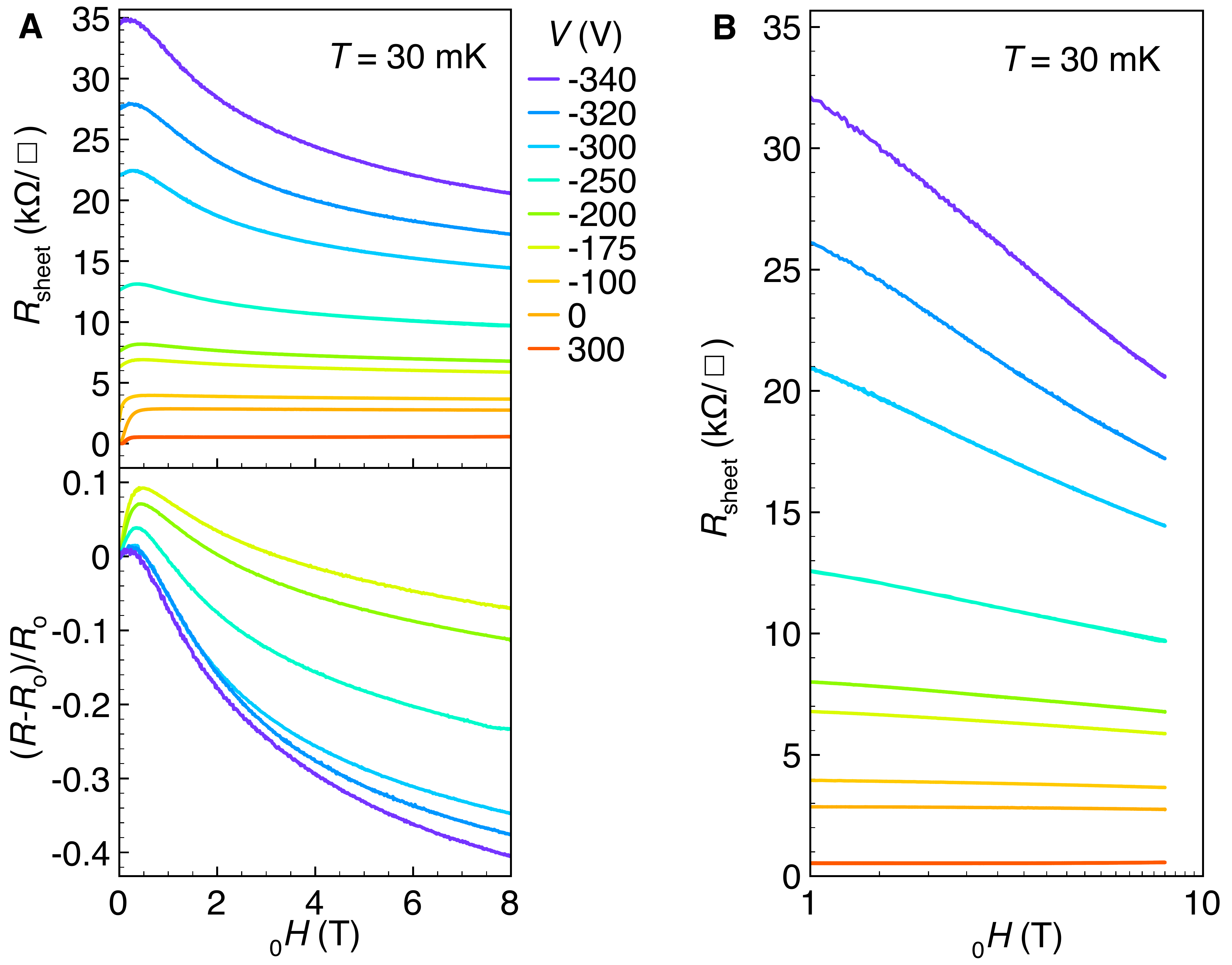}
\caption{\label{fig:mr} Field effect modulation of the magnetotransport properties. (A) Sheet resistance (top panel) and magnetoresistance (bottom panel), as a function of magnetic field, measured for different gate voltages at 30\,mK. Note the large negative magnetoresistance for the lowest carrier densities. (B) Sheet resistance as a function of magnetic field, measured for different gate voltages at 30\,mK and plotted on a logarithmic magnetic field scale.}
\end{figure}

To investigate the quantum critical region in detail we consider a continuous QPT separating a superconducting ground state and an insulating ground state. The control parameter of the phase transition is the variation of the carrier concentration $\delta n_{2D} = n_{2D} - n_{2Dc}$, where $n_{2Dc}$ is the sheet carrier density at the critical point. Fig. \ref{fig:dt}A shows a quasi-linear relationship between the applied gate voltage and
the variation of carrier concentration $\delta n_{2D}\propto \delta
V = V - V_{c}$ in the vicinity of the critical point. Hence we can use the gate voltage as the tuning
parameter for the analysis of the QPT. For a continuous QPT, the quantum
critical region is characterised by a spatial and a temporal correlation length
that diverge respectively as $\xi\propto (\delta n_{2D})^{-\bar{\nu}}$, and $\xi_{\tau}\propto
(\delta n_{2D})^{-\nu_{\tau}}$. The quantum dynamic critical
exponent is defined through the ratio $z=\nu_{\tau}/\bar{\nu}$. According to the scaling theory of quantum critical phenomena \cite{schneider,bookpsc,polonia} the phase transition line
$T_{BKT}(\delta V)$ presented in Fig. \ref{fig:vgtbkt} is expected to scale, in the
vicinity of the quantum critical point (QCP), as
\begin{equation}\label{eq:sca}
T_{BKT}\propto(\delta n_{2D})^{z\bar{\nu}}\propto(\delta V)^{z\bar{\nu}}
\end{equation}
In Fig. \ref{fig:vgtbkt} we observe that the approach to quantum criticality is
well described by $z\bar{\nu} = 2/3$. We note that the product $z\bar{\nu} = 2/3$
agrees with the results obtained in previous experiments of electric
and magnetic field tuned 2D-QSI transition in amorphous bismuth
films \cite{parendo:197004} and Nb$_{0.15}$Si$_{0.85}$ films
\cite{aubin:094521}. $z\bar{\nu} = 2/3$ is compatible with the 3D-XY model possibly indicative of a clean (or weakly disordered) system in which quantum fluctuations dominate ((2+1)D-XY). The voltage range over which critical fluctuations can be observed has however to be determined.
We note that a highly disordered (dirty boson) system \cite{PhysRevLett.60.208} with localised Cooper pairs in the insulating phase would yield to $z\bar{\nu} = 1$. This latter value would agree with the
critical behaviour observed in NdBa$_2$Cu$_3$O$_{7-\delta}$ thin
films \cite{matthey:057002}. Although the dirty boson interpretation
is not incompatible with the data, we notice that in most systems dominated by disorder the magnetoresistance is expected to be positive, at variance with our observations (see below) \footnote{A detailed analysis of the S-I transition will be presented elsewhere.}.

Of particular relevance for the understanding of the nature of this electronic system is the insulating region of the phase diagram. In the accessible temperature range the variation of the conductance can be explained by weak localisation \cite{PhysRevB.57.14440}. Fig.\,\ref{fig:mr} depicts the dependence of the sheet resistance in a perpendicular magnetic field measured at 30\,mK. We observe a large negative
magnetoresistance that increases as we move more deeply into the
insulating phase, reaching more than $-40$\% at 8\,T for the lowest
measured carrier concentration. At this concentration the zero field sheet resistance is $\sim35$\,k$\Omega/\square$. Above $\sim$1\,T the resistance decreases logarithmically with increasing magnetic field (Fig.\,\ref{fig:mr}B) as expected if weak localisation is governing the magnetotransport properties. We note that our data are in partial agreement with the results presented in \cite{Brinkman:2007zr}, where a negative magnetoresistance is reported. However, no hysteresis in magnetoresistance has been detected in our experiments.

\subsection{Methods}
Conducting interfaces were  prepared by depositing LaAlO$_{3}$ films,
with thickness $\geq$ 4 unit cells (u.c.), onto TiO$_{2}$ terminated (001) SrTiO$_{3}$
single crystals. The films were
grown by pulsed laser deposition at $\sim 800$\,$^{\circ}$C in $\sim 1\times
10^{-4}$ mbar of O$_{2}$ with a repetition rate of 1 Hz. The fluence
of the laser pulses was 0.6 J/cm$^{2}$. The film's growth was monitored \textit{in
situ} using reflection high energy electron diffraction which
allowed the thickness to be controlled with sub-unit-cell precision \cite{rijnders:1888}. After growth, each sample was annealed in 200 mbar of  O$_{2}$ at about
600\,$^{\circ}$C for one hour and cooled to room temperature in
the same oxygen pressure.
Samples with different thickness of LaAlO$_{3}$ were systematically characterised using atomic force microscopy and X-Ray
diffraction. The samples were then patterned
in a geometry suitable for 4-point transport measurements. The
patterning technique is based on the discovery that only regions
covered by a LaAlO$_{3}$ layer with thickness $\geq$ 4\,u.c. are conducting \cite{S.Thiel09292006}. The LaAlO$_{3}$ thickness was
thus reduced down to about $<2$\,u.c. in specific regions of the
samples, irradiating them with argon ions while protecting transport
channels with a photoresist layer.

\subsection{Acknowledgements}
We thank Thierry Giamarchi, Lara Benfatto and Thilo Kopp for useful discussions and Marco Lopes for his technical assistance. We acknowledge financial support by the Swiss National Science Foundation through the National Centre of Competence in Research ``Materials with Novel Electronic Properties" MaNEP and Division II, by the European Union through the project ``Nanoxide", by the Deutsche Forschungsgemeinschaft through the SFB484, and by the European Science Foundation through the program ``Thin Films for Novel Oxide Devices".

\end{document}